\documentclass[12pt]{article}
\begin{document}

\begin{center}
{\bf QUANTUM MECHANICS OF DIRAC PARTICLE BEAM TRANSPORT THROUGH OPTICAL
ELEMENTS WITH STRAIGHT AND CURVED AXES}\footnote{To appear in the
Proceedings of the Joint 28th ICFA Advanced Beam Dynamics and Advanced
\& Novel Accelerators Workshop on QUANTUM ASPECTS OF BEAM PHYSICS and Other
Critical Issues of Beams in Physics and Astrophysics, January 7-11, 2003,
Hiroshima University, Higashi-Hiroshima, Japan, Ed. Pisin Chen (World
Scientific, Singapore)} \\
\end{center}

\begin{center}
R.~JAGANATHAN\footnote{{\it E-mail}: {\tt jagan@imsc.res.in}}

{\it The Institute of Mathematical Sciences, \\
4th Cross Road, Central Institutes of Technology Campus, Tharamani \\
Chennai, TN 600113, INDIA} \\
\end{center}

\smallskip

\centerline{\bf Abstract}

\begin{quote}
Classical mechanical treatment of charged particle beam optics is so far
very satisfactory from a practical point of view in applications ranging
from electron microscopy to accelerator technology.  However, it is
desirable to understand the underlying quantum mechanics since the
classical treatment is only an approximation.  Quantum mechanical
treatment of spin-$\frac{1}{2}$ particle beam transport through optical
elements with straight optic axes, based on the proper equation, namely,
the Dirac equation, has been developed already to some extent.  In such a
theory the orbital and spin motions are treated within a unified
framework.  Here, after a brief review of the Dirac spinor beam optics for
systems with straight optic axes it is outlined how the application of
the formalism of general relativity leads to the use of the Dirac equation
for getting a quantum theory of spin-$\frac{1}{2}$ beam transport through
optical elements with curved axes.
\end{quote}

\medskip

\noindent
{\bf 1. Introduction} \\

\noindent
It is surprising that the development of a quantum theory of electron
beam optics based on the proper equation, namely, the Dirac equation, has
only a recent origin \cite{j1,j2,eo}.  The theory of charged particle beam
optics, currently used in the design and operation of various beam
devices, in electron microscopes or accelerators, is largely based on
classical mechanics and classical electrodynamics.  Such a treatment has
indeed been very successful in practice.  Of course, whenever it is
essential, quantum mechanics is used in accelerator physics to understand
those quantum effects which are prominent perturbations to the leading
classical beam dynamics \cite{g}.  The well-known examples are quantum
excitations induced by synchrotron radiation in storage rings, the
Sokolov-Ternov effect of spin polarization induced by synchrotron
radiation, etc.  Recently, attention has been drawn to the limits placed
by quantum mechanics on achievable beam spot sizes in particle
accelerators, and the need for the formulation of quantum beam optics
relevant to such issues \cite{s}.   In the context of electron microscopy
scalar wave mechanics based on the nonrelativistic Schr\"{o}dinger
equation has been the main tool to analyze the image formation and its
characteristics, and the spin aspects are not considered at all \cite{eo}.

In the context of accelerator physics it should be certainly desirable to
have a unified framework based entirely on quantum mechanics to treat the
orbital, spin, radiation, and every aspect of beam dynamics, since the
constituents of the beams concerned are quantum particles.  First, this
should help us understand better the classical theory of beam dynamics.
Secondly, there is already an indication that this is necessary too: it
has been found that quantum corrections can substantially affect the
classical results of tracking for trajectories close to the separatrix,
leading to the suggestion that quantum maps can be useful in finding
quickly the boundaries of nonlinear resonances \cite{hy}.  Thus, a
systematic formalism for obtaining the relevant quantum maps is required.

The aim of this article is present a brief summary of the formalism
quantum beam optics of particles, in particular, Dirac particles, for
treating problems of propagation through optical elements with straight
and curved axes.  \\

\noindent
{\bf 2. Quantum Beam Optics of Particles: An Outline} \\

\noindent
One may consider obtaining the relevant quantum maps for any particle
optical system by quantizing the corresponding classical treatment
directly.  The best way to do this is to use the Lie algebraic formalism
of classical beam dynamics developed particularly in the context of
accelerator physics \cite{lm}.  The question that arises is how to go
beyond and obtain the quantum maps more completely starting {\em ab
initio} with the quantum mechanics of the concerned system since such a
process should lead to other quantum corrections not derivable simply from
the quantization of the classical optical Hamiltonian.  Essentially, one
should obtain a quantum beam optical Hamiltonian $\hat{\mathcal H}$
directly from the original time-dependent Schr\"{o}dinger equation of the
system such that the quantum beam optical Schr\"{o}dinger equation
\begin{equation}
i\hbar\frac{\partial}{\partial z}\psi(\underline{r}_\perp; z)
= \hat{\mathcal H}\psi(\underline{r}_\perp; z)
\label{eq:speq}
\end{equation}
describes the $z$-evolution of the beam wave function
$\psi(\underline{r}_\perp; z)$ where $z$ stands for the coordinate along
the optics axis and $\underline{r}_\perp$ refers to $(x,y)$ coordinates
in the plane perpendicular to the beam at $z$. Since
$\left|\psi\left(\underline{r}_\perp;z\right)\right|^2$ represents the
probability density in the transverse plane at $z$, with
\begin{equation}
\int\int dxdy \left|\psi\left(\underline{r}_\perp;z\right)\right|^2 = 1,
\end{equation}
the average of any observable $\hat{\mathcal O}$ at $z$ is
\begin{eqnarray}
\langle\hat{\mathcal O}\rangle(z)
  & = & \langle\psi(z)|\hat{\mathcal O}|\psi(z)\rangle \nonumber \\
  & = & \int\int dxdy \psi^*(z)\hat{\mathcal O}\psi(z).
\end{eqnarray}
We can write the formal solution of Eq.~(\ref{eq:speq}) as
\begin{eqnarray}
|\psi(z_f)\rangle & = & \hat{U}_{fi}|\psi(z_i)\rangle, \nonumber \\
\hat{U}_{fi} & = & \wp\left\{\exp\left(-\frac{i}{\hbar}
                   \int_{z_i}^{z_f}dz\hat{\mathcal H}(z)\right)\right\},
\end{eqnarray}
where $i$ and $f$ refer to some initial and final transverse planes
situated at $z_i$ and $z_f$, respectively, along the beam axis and
$\wp{\,}$ indicates the path (or $z$) ordering of the exponential.  Then,
the required quantum maps are given by
\begin{eqnarray}
\langle\hat{\mathcal O}\rangle_f
  & = & \langle\psi(z_f)|\hat{\mathcal O}|\psi(z_f)\rangle \nonumber \\
  & = & \langle\psi(z_i)|\hat{U}_{fi}^\dag\hat{\mathcal O}
        \hat{U}_{fi}|\psi(z_i)\rangle \nonumber \\
  & = & \langle\hat{U}_{fi}^\dag\hat{\mathcal O}\hat{U}_{fi}\rangle_i\,.
\label{eq:qmap}
\end{eqnarray}

As an example of the above formalism let us consider a kick in the
$xz$-plane by a thin sextupole represented by the classical phase-space
map
\begin{eqnarray}
x_f & = & x_i, \nonumber \\
p_f & = & p_i + a x_i^2.
\end{eqnarray}
This would correspond to $\hat{U}_{fi}$ $=$ $\exp(\frac{a}{3}\hat{x}^3)$
and following Eq.~(\ref{eq:qmap}) the quantum maps for the averages become
\begin{eqnarray}
\langle\hat{x}\rangle_f & = & \langle\hat{x}\rangle_i, \nonumber \\
\langle\hat{p}\rangle_f & = & \langle\hat{p}\rangle_i +
                              a\langle\hat{x}^2\rangle_i \nonumber \\
                        & = & \langle\hat{p}\rangle_i +
                              a\langle\hat{x}\rangle_i^2
                              + a\langle(\hat{x} -
                              \langle\hat{x}\rangle)^2\rangle_i\,.
\label{eq:xpqmap}
\end{eqnarray}
Now, we can consider the expectation values, such as
$\langle\hat{x}\rangle$ and $\langle\hat{p}\rangle$, as corresponding to
their classical values {\it \`{a} la} Ehrenfest.  Then, as the above
simple example shows, generally, the leading quantum effects on the
classical beam optics can be expected to be due to the uncertainties in
the initial conditions like the term
$a\langle(\hat{x}-\langle\hat{x}\rangle)^2\rangle_i$ in
Eq.~(\ref{eq:xpqmap}).  Such leading quantum corrections involve the
Planck constant $\hbar$ not explicitly but only through the uncertainty
principle which controls the minimum limits for the initial conditions
as has been already pointed out \cite{hy}.  This has been realized earlier
also, particularly in the context of electron microscopy \cite{j3}.

The above theory is only a single-particle theory.  To include the
multiparticle effects, it might be profitable to be guided by models
such as the thermal wave model and the stochastic collective dynamical
model developed for treating the beam phenomenologically as a
quasiclassical many-body system \cite{sl1,sl2}. \\

\noindent
{\bf 3. Quantum Beam Optics of Dirac Particles: Optical Elements
with Straight Axes} \\

\noindent
The proper study of spin-$\frac{1}{2}$ particle beam transport should be
based on the Dirac equation if one wants to treat all the aspects of beam
optics including spin evolution and spin-orbit interaction.  Let us
consider the particle to have mass $m$, electric charge $q$ and an
anomalous magnetic moment $\mu_a$.  It should be noted that the
electromagnetic fields of the optical systems are time-independent.  In
this case one can start with the time-independent equation for the
$4$-component Dirac spinor $\psi\left(\underline{r}_\perp; z\right)$
\begin{eqnarray}
\hat{H}\psi\left(\underline{r}_\perp; z\right)
  & = & E\psi\left(\underline{r}_\perp; z\right), \nonumber \\
\hat{H} & = & \beta mc^2+q\hat{\phi}
  +c\underline{\alpha}_\perp\cdot\underline{\hat{\pi}}_\perp \nonumber \\
  &   & \ + c\alpha_z\left(-i\hbar\frac{\partial}{\partial z}
          - q\hat{A}_z\right)
          - \mu_a\beta\underline{\Sigma}\cdot\underline{B},
\label{eq:tide}
\end{eqnarray}
including the Pauli term to take into account the anomalous magnetic
moment.  Here, we are using the standard notations as is clear from the
context.

Let us assume that we are interested in studying the transport of a
monoenergetic quasiparaxial particle beam through an optical element which
has a straight optic axis along the cartesian $z$-direction.  If $p$
is the design momentum of the beam the energy of a single particle of
the beam is given by $E$ $=$ $\sqrt{m^2c^4+c^2p^2}$.  Further, the
quasiparaxial beam propagating along the $z$-direction should have
$|\underline{p}_\perp|$ $\ll$ $|\underline{p}|$ $=$ $p$ and $p_z$ $>$ $0$.
Then, actually Eq.~(\ref{eq:tide}) has the ideal structure (compare
Eq.~(\ref{eq:speq})) for our purpose since it is already linear in
$\frac{\partial}{\partial z}$.  So, one can readily rearrange the terms in
it to get the desired form of Eq.~(\ref{eq:speq}).  However, it is
difficult to work directly with such an equation since there are problems
associated with the interpretation of the results using the traditional
Schr\"{o}dinger position operator.  In the standard theory of relativistic
quantum mechanics the Foldy-Wouthuysen (FW) transformation technique is
used to reduce the Dirac Hamiltonian to a form suitable for direct
interpretation in terms of the nonrelativistic part and a series of
relativistic corrections.  The FW technique was used originally by
Derbenev and Kondratenko to get their Hamiltonian for radiation
calculations.  This theory has been reviewed and used to suggest a quantum
formulation of Dirac particle beam physics, particularly for polarized
beams, in terms of machine coordinates, observables, and the Wigner
function \cite{hb}.

In an independent and different approach an FW-like technique has been
used to develop a systematic formalism of Dirac particle beam optics in
which the aim has been to expand the Dirac Hamiltonian as a series of
paraxial and nonparaxial approximations \cite{j1,j2,j3,j4}.  This leads to
the reduction of the original $4$-component Dirac spinor to an effective
$2$-component spinor
\begin{equation}
\psi^a\left(\underline{r}_\perp; z\right)
  = \left(\begin{array}{c}
    \psi^a_1\left(\underline{r}_\perp; z\right) \\
    \psi^a_2\left(\underline{r}_\perp; z\right)
    \end{array}\right)
\end{equation}
which satisfies an accelerator optical Schr\"{o}dinger equation
\begin{equation}
i\hbar\frac{\partial}{\partial z}\psi^a\left(\underline{r}_\perp; z\right)
  = \hat{\mathcal H}\psi^a\left(\underline{r}_\perp; z\right).
\label{eq:aose}
\end{equation}
It should be noted that the $2$-component $\psi^a$ is an accelerator
optical approximation of the original $4$-component Dirac spinor, valid
for any value of the design momentum $p$ from nonrelativistic to extreme
relativistic region.

As an example, consider the ideal normal magnetic quadrupole lens
comprising of the magnetic field
\begin{equation}
\underline{B} = (-Gy,\,-Gx,\,0),
\end{equation}
associated with the vector potential
\begin{equation}
\underline{A} = (0,\,0,\,\frac{1}{2}G(x^2-y^2)),
\end{equation}
where $G$ is assumed to be a constant in the lens region and zero outside.
The corresponding quantum accelerator optical Hamiltonian reads
\begin{eqnarray}
\hat{\mathcal H} & \approx &
    \frac{1}{2p}\left(\hat{p}_x^2+\hat{p}_y^2\right)
   -\frac{1}{2}qG\left(\hat{x}^2-\hat{y}^2\right)
   +\frac{1}{8p^3}\left(\hat{p}_x^2+\hat{p}_y^2\right)^2 \nonumber \\
   &    & \ +\frac{q^2G^2\hbar^2}{8p^3}\left(\hat{x}^2+\hat{y}^2\right)
            +\frac{(q+\gamma\epsilon)G}{p}\left(\hat{x}S_y
            +\hat{y}S_x\right),
\label{eq:dqh}
\end{eqnarray}
where $\gamma$ $=$ $E/mc^2$, $\epsilon$ $=$ $2m\mu_a/\hbar$ and
$\underline{S}$ $=$ $\frac{\hbar}{2}\underline{\sigma}$ represents
the spin of the particle defined with reference to its instantaneous
rest frame.  It is to be noted that this quantum accelerator optical
Hamiltonian $\hat{\mathcal H}$ contains all the terms corresponding to the
classical theory plus the $\hbar$-dependent quantum correction terms.
Using the formalism outlined in the previous section it can be shown that
the first two `classical' paraxial terms of the above $\hat{\mathcal H}$
account for the linear phase-space transfer map corresponding to the
focusing (defocusing) action in the $yz$-plane and defocusing (focusing)
action in the $xz$-plane when $G$ $>$ $0$ ($G$ $<$ $0$).  The last
spin-dependent term accounts for the Stern-Gerlach kicks in the
transverse phase-space and the Thomas-Bargmann-Michel-Telegdi spin
evolution \cite{j4}.

The following interesting aspect of quantum beam optics should be
mentioned.  In the case of a spin-$0$ particle also one can derive the
quantum beam optical Hamiltonian $\hat{\mathcal H}$ starting from the
Schr\"{o}dinger-Klein-Gordon equation \cite{j3}.  It would also contain
all the terms corresponding to the classical theory plus the quantum
correction terms.  But, these quantum correction terms are not identical
to the quantum correction terms in the Dirac case.  Thus, besides in the
$\hbar$-dependent effects of spin on the orbital quantum map ({\em e.g.},
the last term in Eq.~(\ref{eq:dqh})), even in the spin-independent quantum
corrections the Dirac particle has its own signature different from that
of a spin-$0$ particle \cite{j5}.  \\

\noindent
{\bf 4. Quantum Beam Optics of Dirac Particles: Optical Elements with
Curved Axes} \\

\noindent
For studying the propagation of spin-$\frac{1}{2}$ particle beams through
optical elements with curved axes it is natural to start with the Dirac
equation written in curvilinear coordinates adapted to the geometry of the
system.  Let us make the $z$-axis coincide with the space curve representing
the optic axis of the system, or the ideal design orbit. Let the transverse,
or off-axis, coordinates $(x,y)$ at each $z$ be defined in such a way that
the spatial arc element $ds$ is given by
\begin{equation}
ds^2 = dx^2+dy^2+\zeta^2dz^2, \qquad
\zeta = (1+\underline{K}_\perp\cdot\underline{r}_\perp),
\end{equation}
where $K_x(z)$ and $K_y(z)$ are the curvature components at $z$.

Now, we have to start with the Dirac equation written in a generally
covariant form.  To this end, let us use the formalism of general
relativity \cite{j2,wls,bw}.  Here, for the sake of simplicity let us
drop the anomalous magnetic moment term.  Then the generally covariant
form of the time-dependent Dirac equation becomes
\begin{eqnarray}
i\hbar\frac{\partial\Psi}{\partial t} & = & \hat{H}\Psi, \nonumber \\
\hat{H} & = & \beta mc^2+q\hat{\phi}
  +c\underline{\alpha}_\perp\cdot\underline{\hat{\pi}}_\perp
  + \frac{c}{\zeta}\alpha_z\left(-i\hbar\frac{\partial}{\partial z}
          - \zeta q\hat{A}_z-\Gamma_z\right), \nonumber \\
\Gamma_z & = & K_xS_y-K_yS_x.
\label{eq:tddeca}
\end{eqnarray}
Further, it should be noted that
\begin{eqnarray}
B_x & = & \frac{1}{\zeta}\left(\frac{\partial(\zeta A_z)}{\partial y}
          -\frac{\partial A_y}{\partial z}\right), \nonumber \\
B_y & = & \frac{1}{\zeta}\left(\frac{\partial A_x}{\partial z}
          -\frac{\partial(\zeta A_z)}{\partial x}\right), \nonumber \\
B_z & = & \left(\frac{\partial A_y}{\partial x}
          -\frac{\partial A_x}{\partial y}\right).
\end{eqnarray}

For a monoenergetic beam with particle energy $E$
\begin{equation}
\Psi(\underline{r},t) = \psi\left(\underline{r}_\perp;z\right)
                        \exp(-iEt/\hbar)
\end{equation}
and $\psi\left(\underline{r}_\perp;z\right)$ satisfies the
time-independent equation
\begin{equation}
\hat{H}\psi\left(\underline{r}_\perp;z\right)
  = E\psi\left(\underline{r}_\perp;z\right)
\label{eq:tideca}
\end{equation}
where $\hat{H}$ is the same as in Eq.~(\ref{eq:tddeca}).  We should now
cast Eq.~(\ref{eq:tideca}) in the form of Eq.~(\ref{eq:speq}) so that the
corresponding beam optical Hamiltonian $\hat{\mathcal H}$ can be derived
and the formalism of Sec.2 can be applied for obtaining the transfer maps
for the quantum averages.  It should be noted that the quantum operators
for the transverse position ($\underline{\hat{r}}_\perp$) and momentum
($\underline{\hat{p}}_\perp$), and spin ($\underline{S}$), are unaltered.
The method of deriving $\hat{\mathcal H}$ proceeds in the same way as for
systems with straight optic axes: a series of FW-like transformations are
to be applied to Eq.~(\ref{eq:tideca}) up to any desired order of accuracy
so that finally a $2$-component equation like Eq.~(\ref{eq:aose}) is
obtained \cite{j2}.  In general, for a magnetic system we get, up to the
first order, or paraxial, approximation,
\begin{equation}
\hat{\mathcal H} = -\zeta p -q\zeta\hat{A}_z +
                    \frac{\zeta}{2p}\hat{\pi}_\perp^2.
\label{eq:caoh}
\end{equation}
For a closed orbit in the $xz$-plane, with no torsion, writing $\zeta$
$=$ $1+\frac{x}{\rho}$, it is clear that $\hat{\mathcal H}$ of
Eq.~(\ref{eq:caoh}) corresponds to the well known Hamiltonian of
classical accelerator optics \cite{l}.  To get a more complete form of
$\hat{\mathcal H}$ including the spin terms and other $\hbar$-dependent
quantum corrections one has to carry out the FW-like transformations
to higher orders.  \\

\noindent
{\bf 5. Concluding Remarks} \\

\noindent
In summary, it is seen that the quantum theory of transport of particle
beams through optical elements is very simple.  Starting from a beam
optical Schr\"{o}dinger equation the transfer maps for quantum averages of
phase-space and spin variables across an optical element can be computed
by a straightforward procedure.  To this end, one has to obtain the
appropriate quantum beam optical Hamiltonian starting from the
corresponding time-dependent Schr\"{o}dinger equation of the system.
As example, quantum theory of propagation of Dirac particle beams through
optical elements with straight and curved optic axes was considered
briefly.  So far, the development of such a theory has not taken into
account multiparticle effects.  Also, such a theory has been developed
only for optical systems.  Taking into account the multiparticle effects
and treating accelerating elements are issues of the theory to be tackled
in future.  \\

\noindent
{\bf Acknowledgments} \\

\noindent
I am very much thankful to Prof. Pisin Chen and Prof. Atsushi Ogata for
the warm hospitality.  I would also like to thank our Institute, and the
director Prof. R. Balasubramanian, for the financial support for travel
which made my participation in this workshop possible.

\end{document}